\documentclass[twocolumn, superscriptaddress, amsmath, amssymb, aps]{revtex4-2}

\usepackage{graphicx}% Include figure files
\usepackage{dcolumn}% Align table columns on decimal point
\usepackage{bm}% bold math
\usepackage{xcolor}
\usepackage{mathtools}
\usepackage{physics}
\usepackage[normalem]{ulem}
\usepackage{soul}
\usepackage{braket}
\usepackage{appendix}
\usepackage{xcolor}   
\usepackage{xr}
\usepackage{array} % for defining new column types
\usepackage{hyperref} % the last package to be imported
\hypersetup{colorlinks=true, citecolor=blue}

\newcommand{\rev}[1]{{\color{black} #1}} 

\newcolumntype{P}[1]{>{\centering\arraybackslash}p{#1}}

\begin{document}
%TC:ignore
\title{Measurement and Optimal Targeting of a Hidden Scatterer in a Complex Environment Utilizing Fisher Information}

\author{Nadav Shaibe}
\email{Corresponding author: nshaibe@umd.edu}
 \affiliation{Maryland Quantum Materials Center, Department of Physics, University of Maryland, College Park, Maryland 20742-4111, USA}

\author{Jared Erb}
 \affiliation{Maryland Quantum Materials Center, Department of Physics, University of Maryland, College Park, Maryland 20742-4111, USA}
 
\author{Steven M. Anlage}
 \affiliation{Maryland Quantum Materials Center, Department of Physics, University of Maryland, College Park, Maryland 20742-4111, USA}

\date{\today}

\begin{abstract}

A complex, non-Hermitian scattering system with a high degree of multiple scattering and interference is often treated as a black box, described simply by the relationship between a set of incoming and outgoing waves of a given frequency or energy. The scattering matrix $S$ that describes the system is a non-unique, generally sub-unitary matrix that reveals very little about the microscopic processes that are responsible for the observed scattering. We form the Fisher information operator $F_x$, a Hermitian matrix, utilizing a derivative of $S$ with respect to the value of some varying parameter $x$ of the system, associated with a localized perturbation, and experimentally demonstrate that the principal eigenvector of $F_x$ can be used to quantitatively measure  changes in the value of parameter $x$ using only information from the scattering matrix. \rev{The parameter changes are compared to the Cram\'{e}r-Rao bound and found to be close to the precision limit}. We propose a ``discrete feedback loop'' protocol enabled by the knowledge we gain from the Fisher information operator,  that repeatedly determines the counter perturbation necessary to return a varying parameter to some fixed benchmark value. A further application of the Fisher information operator for energy focusing that utilizes the principal eigenvector excitation is demonstrated through compelling indirect evidence of targeting within a complex system. These methods are experimentally demonstrated to work even in the presence of time-reversal symmetry breaking due to absorption and/or loss of scattering reciprocity.

\end{abstract}

\maketitle
%\newpage

%\section{Introduction}
%TC:endignore

\section{Introduction}\label{SEC_Intro}

% \textit{Introduction}.---
Real-world systems are often disordered or complex, with an exact structure that might be unknown or even hidden. Waves incident on such a complex medium or inside a complex cavity experience random scattering and interference, and after just a few scattering events can be challenging to track in experiment or simulation. Instead, we borrow the ideas and formalism developed in mesoscopic physics and describe these systems using scattering theory \cite{Gopar1996,Beenakker1997,Alhassid2000,Kottos2005,Li2017,Rotter2017,Fyodorov2017}. The  $M\times M$ complex scattering matrix relates outgoing waves to incoming waves by $|E_\mathrm{out}\rangle=S|E_\mathrm{in}\rangle$, $M$ being the number of scattering channels connected to a closed system. We note that the scattering matrix can be related to the effective Hamiltonian of the closed system through the Heidelberg approach \cite{VERBAARSCHOT1998,SOKOLOV1989,Fyodorov1997bTRI,Kuhl2013,Schomerus2015}.

The scattering matrix, or similar operators such as the transfer matrix that describe reflection and transmission, is a very effective description for the behavior of a system as a whole and has been used successfully in microwave cavities and networks \cite{Hul2012,Gradoni2014,Lawniczak2023,Lei2024,erb2025,shaibe2025}, quantum billiards \cite{Stockmann1990,Fyodorov2004,Dietz2015}, acoustics \cite{Auregan16,Shi2016,Achilleos2017}, optics \cite{Lin2011,Popoff2014,Huang2020,Genack2024}, and beyond. Despite its utility and experimental accessibility, $S$ tells us virtually nothing specific about what is going on within the system. For this reason, there has recently been a push to use the scattering matrix and its elements to develop estimates of the internal happenings, such as deriving bounds on the internal field amplitudes \cite{Sounas2017,Ma2025} or constraints on transport phenomena \cite{Guo2024}. In this paper, our particular interest is to use the scattering matrix to learn about a parameter $x$ that is hidden within the system and not directly accessible. In general, the outgoing wave $|E_\mathrm{out}(\hat{x})\rangle$ at $\hat{x}=x_0+\Delta x$, where $\Delta x$ is a small perturbation of the parameter from its benchmark value $x_0$, can be described in terms of $|E_\mathrm{out}(x)\rangle$ by expanding linearly as follows:
\begin{align}
        |E_\mathrm{out}(\hat{x})\rangle &=S(\hat{x})|E_\mathrm{in}\rangle \nonumber \\
        &=S(x_0)|E_\mathrm{in}\rangle+\Delta x\frac{\partial S}{\partial x}\Big|_{x=x_0}|E_\mathrm{in}\rangle \nonumber\\
        &=|E_\mathrm{out}(x_0)\rangle+\Delta x|\partial_xE_\mathrm{out}(x_0)\rangle \nonumber
\end{align}
where $|\partial_xE_\mathrm{out}(x_0)\rangle$ is the derivative of $|E_\mathrm{out}\rangle$ with respect to $x$ at $x_0$. The \rev{unbiased linear} estimator of $\Delta x$ is then \cite{Kay1993,Bouchet2021}:
\begin{equation}
    \Delta x = \frac{\mathrm{Re}\Big[\langle\partial_x E_\mathrm{out}(x_0)|E_\mathrm{out}(\hat{x})\rangle-\langle\partial_x E_\mathrm{out}(x_0)|E_\mathrm{out}(x_0)\rangle\Big]}
    {\langle\partial_x E_\mathrm{out}(x_0)|\partial_x E_\mathrm{out}(x_0)\rangle}. \label{EQN_EST}
\end{equation}

However, not every $|E_\mathrm{in}\rangle$ can be used to get an accurate estimate of $\Delta x$. What is required is an incoming wave that creates system excitations which interact with the parameter $x$ and carry its imprint via $|E_\mathrm{out}\rangle$. The amount of information an observable ($|E_\mathrm{out}\rangle$ in this case) carries about an unknown parameter is quantified by the Fisher information \cite{Kay1993}. Fisher information has been used in the localization of tiny objects such as single molecules or nanowires \cite{Chao2016,Yuan2019,Bouchet2020,Grant2025,Graaff2025,Weimar2026}, improving quantum metrology \cite{Giovannetti2011,Tsang2011,Tsant2016,Yang2017,Wiersig2026,Liu2026}, and for optimizing wave propagation through complex systems to maximize desired information at the receivers \cite{Bouchet2021,Horodynski2021,Bouchet2023,Hupfl2024}.

In this work we use the Fisher information operator (FIO) matrix defined by Bouchet, Rotter, and Mosk in Ref.~\cite{Bouchet2021}, which is written in terms of the scattering matrix as:
\begin{equation}
    F_x= \left(\frac{\partial S}{\partial x}\right)^\dagger\frac{\partial S}{\partial x}. \label{EQN_FISH}
\end{equation}
Note that if the scattering matrix is not feasibly or physically measurable in its entirety, Eq.~\ref{EQN_FISH} can also be formed using sub-blocks of $S$, such as the reflection $R$ or transmission $T$ matrices. $F_x$ is a Hermitian operator with the simple property that its eigenvectors $|E_m\rangle$ ($m\in[1,\dots,M]$) are ranked by their associated eigenvalues $\lambda_m$ corresponding to how much information about $x$ is returned by $|E_\mathrm{out,m}\rangle=S|E_m\rangle$. The FIO eigenvalues $\lambda_m$ have the units of $x$ inverse squared and are bounded below by $0$ but have no upper bound as far as we know. Our preliminary statistical investigations suggest there is a suppression of very large $\lambda$ values \rev{that appears to be} exponential, \rev{more details are provided in Appendix~\ref{APP_STATS}}. Because $F_x$ is Hermitian, its eigenvectors form a complete orthogonal basis so all possible $|E_\mathrm{in}\rangle$ can be written as a linear combination of the $|E_m\rangle$'s. It naturally follows that the principal eigenvector $|E_1\rangle$ associated with the largest eigenvalue $\lambda_1$ is the optimal choice of $|E_\mathrm{in}\rangle$ for investigating the hidden parameter $x$.

In this paper we construct $F_x$ for a complex microwave scattering system and use it for two distinct purposes. First,  we show experimentally that the principal eigenvector of $F_x$ allows small changes in the value of a hidden parameter $x$ to be estimated with far greater accuracy than any other excitation, and we use this to implement a ``discrete feedback loop" that actively stabilizes the system against parameter drift. Second, we indirectly show that this same principal eigenvector also optimally focuses energy to the physical location of the localized parametric change. Our approach connects existing wavefront-shaping methods for focusing energy inside disordered and reverberant media \cite{Prada1994,Mosk2012,Ambichl2017,Cao2022,Goicoechea2025} to perturbed guidestar techniques \cite{Zhou2014,Ma2014,Ruan2017} but has added advantages. The first is that any parameter with a changing value $(x)$ that perturbs the scattering process can be used as the guidestar, regardless of whether that change is of its location, electromagnetic or optical properties, or anything else. Another advantage is that the Fisher information operator approach continues to work when time reversal symmetry is broken, for instance by large amounts of absorption or through the introduction of non-reciprocity due to wave propagation through a magnetized ferrite \cite{PaulSo95,Stock95,Dietz2009,Lawniczak2010}, because it does not rely on the existence of time-reversal invariance.

\section{Determination of Parameter Value Through Scattering Measurements}\label{SEC_MIS}

\subsection{\rev{Measurement Scheme}}

% \textit{Experiments}.---
The first experiment we report was conducted by measuring the $M\times M$ scattering matrix $S$ of a quasi-two-dimensional quarter bowtie microwave billiard \cite{PaulSo95,Gokirmak98} as a function of two parameters, \rev{linear} frequency $\omega$ and the voltage $V$ applied to a single globally-biased varactor diode-loaded metasurface $TM^{1D}_1$ \cite{Sleasman2023,Erb2024} using a four port microwave network analyzer.  A schematic of the billiard with five metasurface segments is shown in Fig.~\ref{FIG_1}(a). By tuning the voltage applied to the metasurface we can change the reflection amplitude and phase of waves that interact with it. Because the system is highly reverberant, the waves will interact with the metasurface multiple times, giving us a strong degree of control over the scattering properties of the system, despite the finite size of the metasurface. The quarter bowtie billiard has a perimeter of $1.574$ meters, and the metasurfaces are all identical, being comprised of 18 unit cell varactor diode-loaded elements arranged in a linear array with a total length of $18.5$ cm. Each metasurface covers $11.75\%$ of the billiard's perimeter.

With the scattering matrix measured, we construct the Fisher information operator $F_V=\big(\frac{\partial S}{\partial V}\big)^\dagger\frac{\partial S}{\partial V}$ at each frequency and metasurface voltage value.  The choice of $\omega_0$ and $V_0$ is arbitrary, so we measure over the \rev{effective operating} domain of the metasurface which is $8-10$ GHz and $0-12$ V.  Once we have decided on the frequency $\omega_{0}$ and voltage $V_{0}$ of interest, and subsequently the input excitation $|E_\mathrm{in}\rangle$, there are two ways we can determine the outgoing waves $|E_\mathrm{out}\rangle$ necessary for Eq.~\ref{EQN_EST}. The simpler but less direct method is using only scattering matrix measurements by taking the mathematical operation $|E_\mathrm{out}(\hat{V})\rangle=S(\hat{V})|E_\mathrm{in}\rangle$.

The more direct method is to change the network analyzer state from standard S-parameter measurement mode to multiple independent source injection mode. Using two external analog signal generators and the two internal microwave sources in the network analyzer, we are able to directly create any complex wavefront $|E_{\mathrm{in}}\rangle$ with up to 4 elements to inject into our system, as well as measure the complex outgoing wavefront $|E_{\mathrm{out}}\rangle$ on four receivers. With either method, we utilize Eq.~\ref{EQN_EST} to  estimate $\Delta V$, the change in the metasurface applied bias voltage. We stress that this is a very general procedure that can be done for any parameter that can be toggled between at least two states that sufficiently perturb the scattering matrix, whatever those states may be.

\subsection{\rev{Verification of Quantitative Information Recovery}}

\begin{figure} [thb]
	\centering
	\includegraphics[width=0.95\columnwidth]{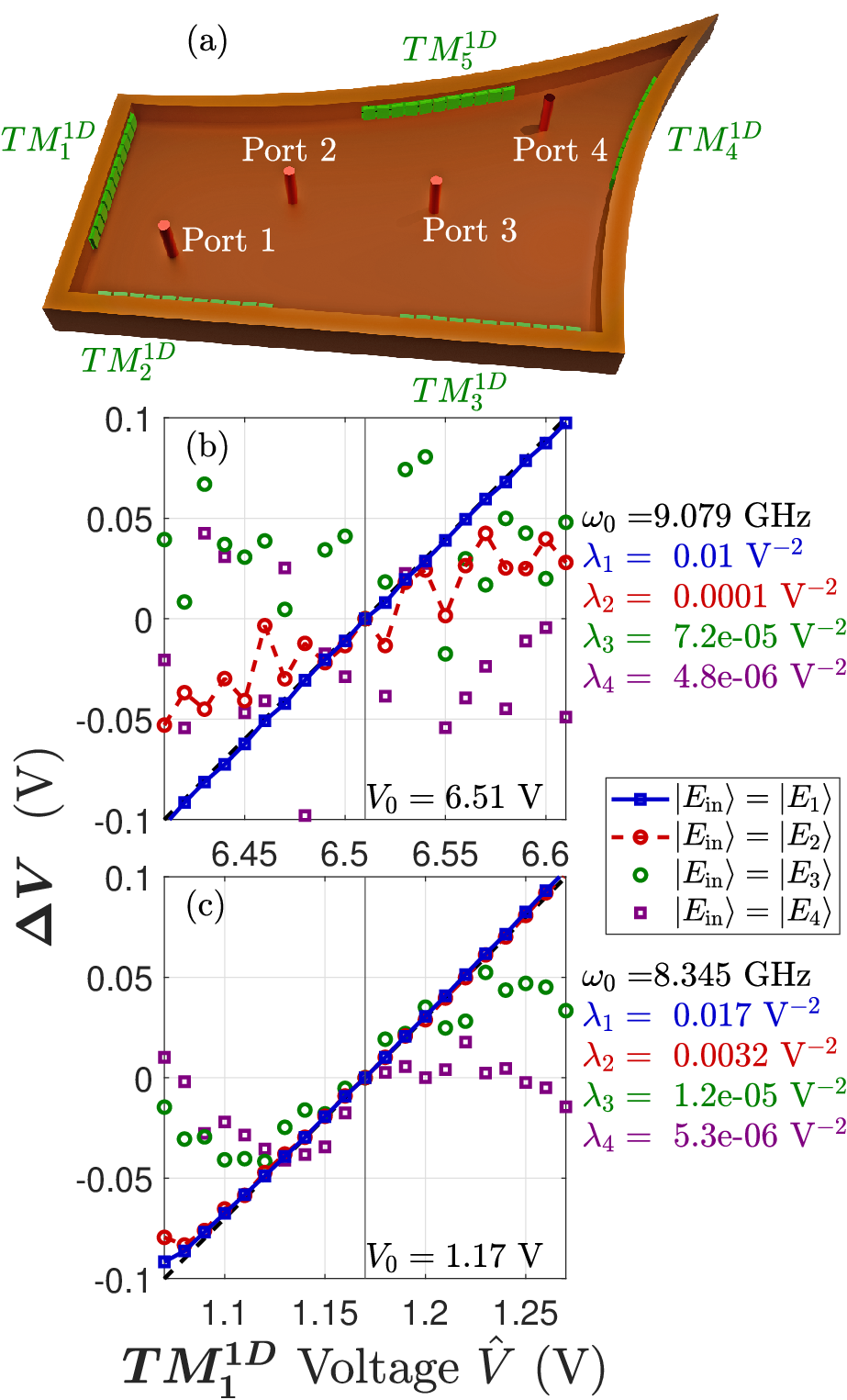}
	\caption{(a) Schematic of quarter bowtie microwave billiard (lid removed) with $M=4$ ports (red cylinders) and five tunable metasurfaces $TM^{1D}_i$, $i\in[1,\dots,5]$ (green strips). (b-c) Estimate of the change of voltage $\Delta V$ applied to $TM^{1D}_1$, calculated from Eq.~\ref{EQN_EST} using the four eigenvectors of $F_V$ for $|E_{\mathrm{in}}\rangle$. Dashed black line depicts true change, which is a straight line with a slope of 1. Associated eigenvalues of $F_V$ and frequency of the measurement are given on the right hand side.}
	\label{FIG_1}
\end{figure}

For Fig.~\ref{FIG_1}, we directly measure the complex outgoing wave $|E_{\mathrm{out}}(\hat{V})\rangle$ while holding $|E_{\mathrm{in}}\rangle$ fixed as one of the eigenvectors of $F_V(\omega_0,V_0)$. An estimate of $\Delta V=\hat{V}-V_0$ is determined using Eq.~\ref{EQN_EST}, and then we repeat the process for the next eigenvector of $F_V$.

Empirically, we have found two classes of quantitative parameter determination. Fig.~\ref{FIG_1}(b) shows a demonstrative example of the first class of cases, which corresponds to when $\lambda_1\geq\lambda_2\times 10^{2}$ in an $M=4$ port system. Using the principal eigenvector $|E_1\rangle$ of $F_V$ as the excitation wave results in a very good quantitative estimate of $\Delta V$ (in Volts), while all other choices of $|E_\mathrm{in}\rangle$ cannot accurately estimate $\Delta V$. Note that the estimated values of $\Delta V$ fall on a line of slope 1 when plotted against the actual voltage values, centered on the arbitrarily chosen value $V_0$, demonstrating the ability to quantitatively measure the change in the hidden parameter through scattering measurements alone.  Panel (c) shows the second class of cases ($\lambda_1\leq\lambda_2\times 10^{2}$), where both the principal eigenvector and the eigenvector of $F_V$ associated with the second largest eigenvalue can both be used to recover good quantitative estimates of the parameter change using Eq.~\ref{EQN_EST}. The two orders of magnitude threshold between the two cases is an empirical rule for this specific billiard and metasurface. It could be that for a different system with a different parameter $x$, the ratio of the largest two eigenvalues in the two cases may be different. 

We have observed both classes of cases even in $M=2$ port systems, the second case resulting in any arbitrary $|E_\mathrm{in}\rangle$ being able to give accurate estimates of $\Delta x$. We have never experimentally found a situation where the third or higher eigenvector of $F_V$ gives a successful estimate. We are currently not aware of any reason this would be forbidden, but each subsequent eigenvalue of $F_V$ gets smaller and smaller so \rev{it becomes} unrealistic to expect to be able to determine anything about a parameter with so little Fisher information. Generically, only the principal eigenvector $|E_1\rangle$ can be relied upon for this purpose. Comparing panels (b) and (c) of Fig.~\ref{FIG_1}, it's clear that the value of the FIO eigenvalue is important for whether its associated eigenvector can be used to accurately estimate $\Delta x$, as in panel (b) $\lambda_2$ is too small, but in panel (c) $\lambda_2$ is large enough that $|E_2\rangle$ returns a good estimate. In our experiments, $|E_1\rangle$ never failed to give an accurate estimate of $\Delta x$, and this is because the metasurface we use is by design highly perturbative to the scattering matrix.

In Appendix \ref{APP_BTRI}, we show how we are able to use the FIO to estimate changes in the length of a bond of an $M=6$ port non-reciprocal and lossy microwave graph, where the length changes are on the order of millimeters and the overall length of the graph is over $3.5$ meters. This demonstrates the use and effectiveness of this method in systems without Lorentz symmetry, and in the case of qualitatively different kinds of parameters.

\subsection{\rev{Comparison with Cram\'{e}r-Rao Bound}}

\rev{Because Eq.~\ref{EQN_EST} is the \textit{minimum variance} unbiased estimator, in the regime of Gaussian noise its variance reaches the Cram\'{e}r-Rao bound \cite{Kay1993,vantrees2013,Bouchet2021,Bouchet2023}. In this section, we compare our results to the Cram\'{e}r-Rao bound by repeatedly estimating the change in the voltage $\Delta V$ applied to the metasurface $TM_1^{1D}$. Further characterization of the noise in our measurements is available in Appendix~\ref{APP_NOISE}.}

\rev{Depending on the type of measurement, we write the Cram\'{e}r-Rao inequality as either:}

\begin{equation}
    \rev{\mathrm{var}(\Delta x)\geq\frac{\sigma^2_{|E_\mathrm{out}\rangle}}{\mathcal{A}^2}\frac{1}{\lambda_1}} \label{EQN_CRInj}
\end{equation}
\rev{when we perform multi-source injection measurements, where $\mathcal{A}^2=\langle E_\mathrm{in}|E_{\mathrm{in}}\rangle$ is the squared amplitude of the input signal and $\sigma^2_{|E_\mathrm{out}\rangle}=\mathrm{var}\left(\sqrt{\langle E_\mathrm{out}|E_{\mathrm{out}}\rangle} \right)$ is the variance of the amplitude of the output signal, or}
\begin{equation}
    \rev{\mathrm{var}(\Delta x)\geq\frac{\sigma^2_{\mathrm{det}(S)}}{\lambda_1}} \label{EQN_CRSmt}
\end{equation}
\rev{when we perform scattering matrix measurements, where $\sigma^2_{\mathrm{det}(S)}$ is the variance of the determinant of the scattering matrix. For both forms, $\lambda_1$ is the principal FIO eigenvalue. }

\begin{figure} [htb]
	\centering
	\includegraphics[width=1\columnwidth]{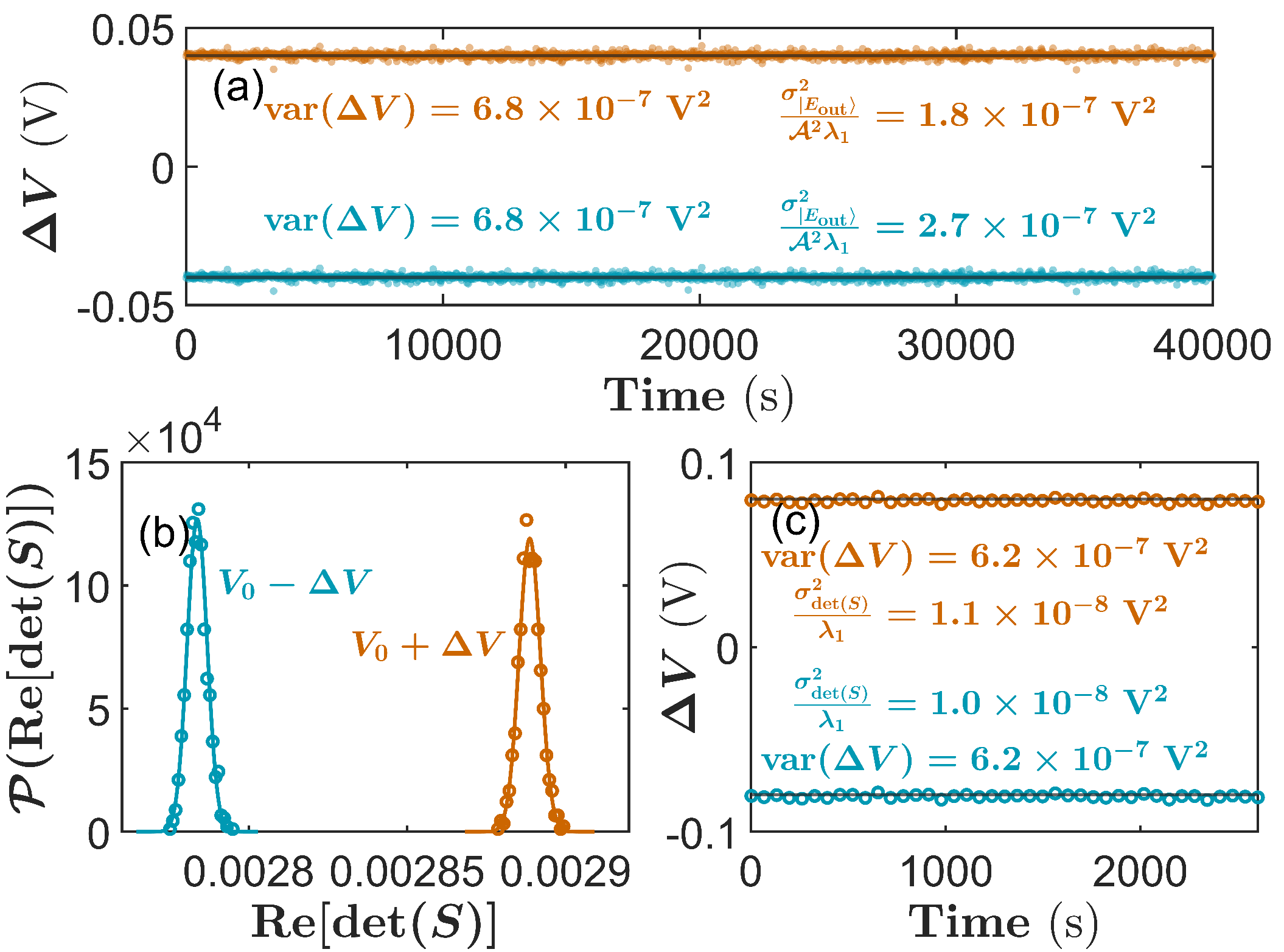}
	\caption{\rev{(a) Estimated $\Delta V$ of metasurface $TM_1^{1D}$ in the $M=2$ port quarter bowtie billiard measured at $\omega_0=9.5~\mathrm{GHz}$, $V_0=3.16~\mathrm{V}$, and true $\Delta V=0.04~\mathrm{V}$ over a period of 11 hours. The simultaneous injection method was used to directly measure $|E_\mathrm{out}(\hat{V})\rangle$ by exciting the system with the principal FIO eigenvector $|E_1\rangle$. The principal FIO eigenvalue for this system is $\lambda_1=0.037~\mathrm{V^{-2}}$. The left and right hand side of the Cram\'{e}r-Rao inequality (Eq.~\ref{EQN_CRInj}) are provided. (b) Probability density functions (PDFs) of the real component of the determinant of the scattering matrix $\mathrm{det}(S)$ of the $M=4$ port quarter bowtie billiard measured at $\omega_0=8.775~\mathrm{GHz}$, $V_0=1.6~\mathrm{V}$, and $\Delta V=0.08~\mathrm{V}$ measured over 1 hour. Circles correspond to experimental data and the solid line shows a Gaussian fit. (c) The estimated $\Delta V$ from the scattering matrix measurement in panel (b), with every 25$^\mathrm{th}$ point shown. The principal FIO eigenvalue for this system is $\lambda_1=0.011~\mathrm{V^{-2}}$. The left and right hand side of the Cram\'{e}r-Rao inequality (Eq.~\ref{EQN_CRSmt}) are provided.}}
	\label{FIG_CR_S}
\end{figure}

\rev{We demonstrate close agreement to the bound defined by both equations in Fig.~\ref{FIG_CR_S}. In panel (a), we show the estimated $\Delta V$ from a direct injection measurement in the $M=2$ port quarter bowtie billiard measured at $\omega_0=9.5~\mathrm{GHz}$, $V_0=3.16~\mathrm{V}$, and $\Delta V=0.04~\mathrm{V}$. We repeatedly excited the system with the principal $F_V$ eigenvector $|E_1\rangle$ for 11 hours while toggling $TM_1^{1D}$, resulting in 1667 measurements at each metasurface voltage. We found that the variance of the estimated $\Delta V$ was $6.8\times10^{-7}~\mathrm{V}^{2}$ for both $\hat{V}=3.12~\mathrm{V}$ and $\hat{V}=3.2~\mathrm{V}$. The input wavefront amplitude $\mathcal{A}$ was held fixed at $0.0703~\mathrm{V}$, but the output amplitudes did have slightly different variances, $3.3\times10^{-11}~\mathrm{V}^2$ at $\hat{V}=3.2~\mathrm{V}$ and $4.9\times10^{-11}~\mathrm{V}^2$ at $\hat{V}=3.12~\mathrm{V}$. The principal FIO eigenvalue is $\lambda_1=0.037~\mathrm{V^{-2}}$, and we find the estimator variances to be in good agreement with the Cram\'{e}r-Rao bound given by Eq.~\ref{EQN_CRInj}.}

% \begin{equation}
%     \rev{\mathrm{var}(\Delta x)=6.8\times10^{-7}~\mathrm{V}^{-2}\geq1.8\times10^{-7}~\mathrm{V}^{-2}=\frac{\sigma^2_{|E_\mathrm{out}\rangle}}{\mathcal{A}^2}\frac{1}{\lambda_1}} \nonumber
% \end{equation}
% \rev{at $\hat{V}=3.2~\mathrm{V}$, and}
% \begin{equation}
%     \rev{\mathrm{var}(\Delta x)=6.8\times10^{-7}~\mathrm{V}^{-2}\geq2.7\times10^{-7}~\mathrm{V}^{-2}=\frac{\sigma^2_{|E_\mathrm{out}\rangle}}{\mathcal{A}^2}\frac{1}{\lambda_1}} \nonumber
% \end{equation}
% \rev{at $\hat{V}=3.12~\mathrm{V}$.}

\rev{In Fig.~\ref{FIG_CR_S}(b), we show probability density functions (PDFs) of $\mathrm{Re}[\mathrm{det}(S)]$ of the $M=4$ port quarter bowtie billiard measured at $\omega_0=8.775~\mathrm{GHz}$, and $\hat{V}=1.6\pm0.08~\mathrm{V}$. The scattering matrices were measured repeatedly for 1 hour, resulting in 5001 measurements at each voltage. The circles correspond to the PDFs of the experimental data while the solid curves are Gaussians with the same means and standard deviations as the data. We arbitrarily chose to show the real component of the determinant of $S$, but the PDFs of $\mathrm{Im}[\mathrm{det}(S)]$ are also Gaussian. Panel (c) has the same type of data as panel (a), but since we did not do direct injections for this measurement, to estimate $\Delta V$ we mathematically operated on the measured scattering matrices with the calculated $F_V$ principal eigenvector. The circles show the estimated change at every $25^\mathrm{th}$ time step. The FIO principal eigenvalue is $\lambda_1=0.011~\mathrm{V}^{-2}$ and we find $\sigma^2_{\mathrm{det}(S)}=1.1\times10^{-8}$ at $\hat{V}=1.68~\mathrm{V}$ and $\sigma^2_{\mathrm{det}(S)}=1.0\times10^{-8}$ at $\hat{V}=1.52~\mathrm{V}$. Once again, even though the method is different, we get the same estimator variance $\mathrm{var}(\Delta V)=6.2\times10^{-7}~\mathrm{V^{2}}$ at both $\hat{V}=1.68~\mathrm{V}$ and $\hat{V}=1.52~\mathrm{V}$. These variances are nearly down to the precision limit defined by Eq.~\ref{EQN_CRSmt}.}

% \rev{ The calculated Crame\'{e}r-Rao bound is lower for this example than the previous, and the variance of our estimator isn't as close. One possibility }

\subsection{\rev{Discrete Feedback Loop}}

Since we can accurately estimate small changes of a parameter, we are able to keep the system stable against perturbations of that parameter. By repeatedly estimating the change from a desired value and applying the inverse perturbation, a ``discrete feedback loop'' can be built. We call it discrete because finite time intervals are required to measure the new state, estimate the parameter change, and apply the appropriate counter change. \rev{This protocol could be useful in situations where direct observation of a fluctuating parameter is unavailable, with the advantage that no additional measurement streams are needed. Simply through the already collected measurements of reflection or transmission, the state of the parameter can be estimated with high accuracy, assuming the system can be prepared such that $\lambda_1$ is large. }

\begin{figure} [thb]
	\centering
\includegraphics[width=1\columnwidth]{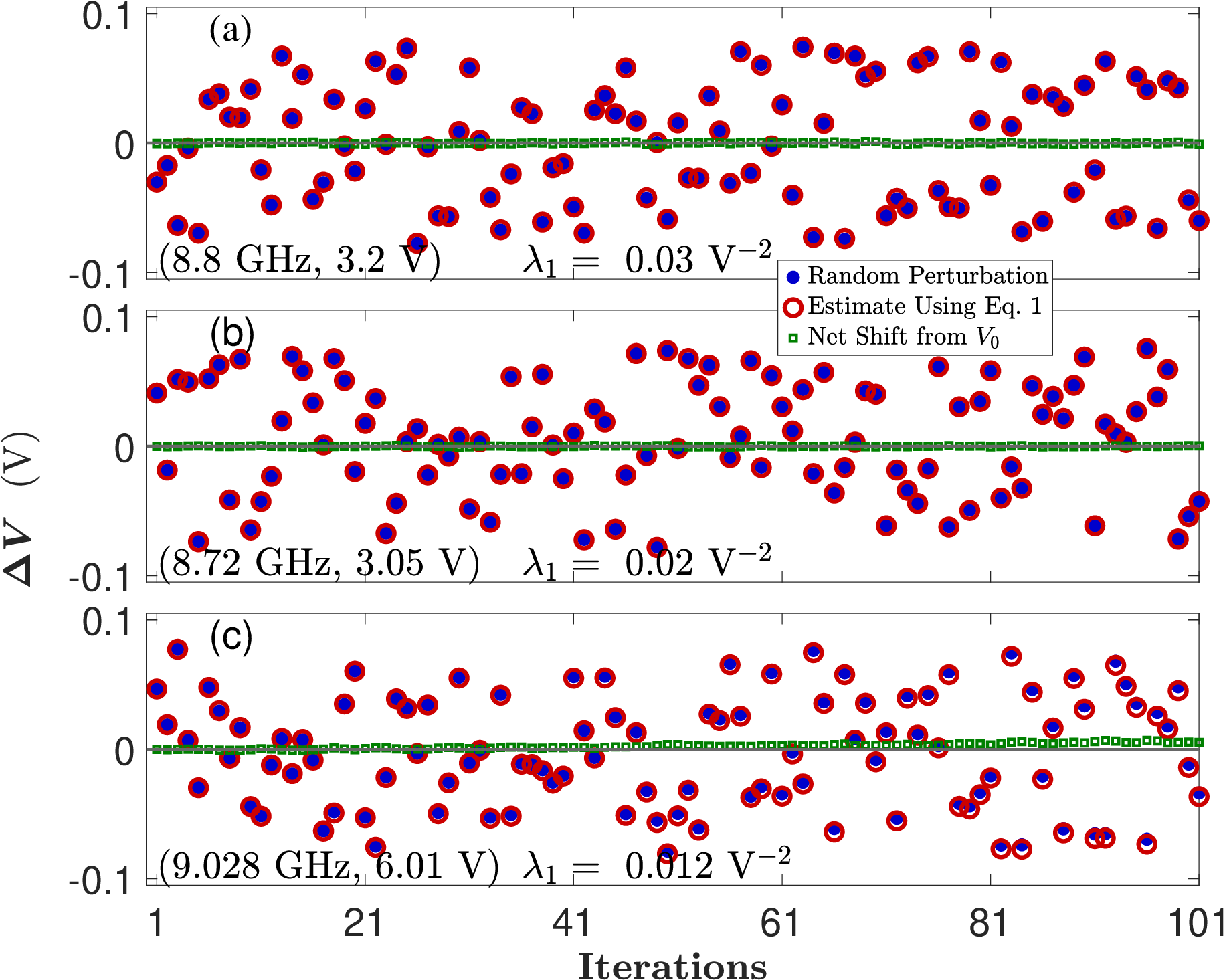}
	\caption{Change in $TM_1^{1D}$ voltage $\Delta V$ across 101 iterations of the ``discrete feedback loop'' process. Blue dots are the known perturbations of each step while the red circles are the estimates for $\Delta V$ using Eq.~\ref{EQN_EST}. Green squares depict the net voltage shift from $V_0$ at each iteration after using the estimate to counter act the perturbation. Panels (a) and (b) are done using scattering matrix measurements while panel (c) is done using simultaneous injection measurements. The ($\omega_0$,$V_0$) for each case is given, along with the principal FIO eigenvalue. The final net shifts in voltage from $V_0$, mean shift, and standard deviation are: (a) $-7.75\times 10^{-4}$ V, $-2.82\times 10^{-5}$ V, $4.18\times 10^{-4}$ V), (b) $4.12\times 10^{-4}$ V, $6.68\times 10^{-6}$ V, $2.83\times 10^{-4}$ V), and (c) $5.46\times 10^{-3}$ V, $2.50\times 10^{-3}$ V, $2.05\times 10^{-3}$ V.}
	\label{FIG_2}
\end{figure}

We implement this stabilizing protocol in the quarter bowtie billiard, showing the results over 101 second-long iterations for three different cases shown in Fig.~\ref{FIG_2}. For all three panels, the blue dots show a randomly generated perturbation of the voltage applied to metasurface $TM_1^{1D}$ in the range of $[-0.08,0.08]$ V, chosen from a uniform distribution. We then estimate $\Delta V$ through changes in the scattering properties of the system using Eq.~\ref{EQN_EST}, depicted by the red circles. The green squares show the difference between the fixed desired voltage $V_0$ and the actual voltage after applying $-\Delta V$. In the first case, shown in panel (a), the billiard has $M=4$ ports, and we are repeatedly measuring the scattering matrix and then calculating  $|E_\mathrm{out}(\hat{V})\rangle$ via $S(\hat{V})|E_\mathrm{in}\rangle$ where $|E_\mathrm{in}\rangle$ is the principal eigenvector of $F_V$. In panel (b), we are also using scattering matrix measurements but with $M=2$ ports, and in panel (c) we simultaneously excite both ports of the two-port billiard and directly measure $|E_\mathrm{out}\rangle$. \rev{Note that in none of these examples do we average multiple measurements together to determine an estimate of $\Delta V$. The FIO is determined from a single finite difference before the random perturbations start, and the estimates of $\Delta V$ are completed faster than our instrument communication time.}

Fig.~\ref{FIG_2}(c) is unique among the three in that after about 25 iterations, the estimates begin to have a systematic bias to be more positive than the actual perturbations, leading to a drift in voltage away from $V_0$. This suggests there is some changing parameter other than the voltage applied to $TM_1^{1D}$. The other four metasurfaces are not being biased with any voltage, and the quarter bowtie billiard used in this measurement is fairly stable on the scale of hundreds of hours (see Sec.~VIII of Supp. Mat. of Ref.~\cite{Erb2026}). The culprits are the two parameters added when doing a simultaneous two port injection: the power ratio and phase difference across the two injection ports. We can consider the scattering matrix as being nearly fixed, but the excitation wave $|E_\mathrm{in}\rangle$ itself is drifting in time. In the four port direct injection case (not shown in Fig.~\ref{FIG_2}), where there are six introduced parameters and we need to use the external microwave sources which are less stable than the network analyzer's internal sources, we were unable to keep the \rev{input wavefront} fixed \rev{to $|E_1\rangle$}. \rev{This is discussed further in Appendix~\ref{APP_NOISE}}.

\section{Fisher Information Enabled Targeting Inside a Complex System}\label{SEC_HIT}

Focusing energy on a specific target inside complex media is a well studied problem, often motivated by applications of imaging medical and biological samples \cite{Prada1994,Vellekoop2007,Popoff2010,Horstmeyer2015}. In disordered media with multiple random scattering events, wavefront shaping techniques have been developed to take advantage of interference effects \cite{Mosk2012,Hougne2016}. There is also the idea of wireless power transfer (WPT), which can be done effectively in free-space using various beamforming architectures \cite{Hui2014,Zeng2016}, but is much more challenging in complex scattering environments. Time reversal methods have been proposed to avoid the limitations of beaming \cite{Lerosey2006,Xu2011,Cangialosi2016}, and more recently a cavity-wavefront shaping technique making use of the Coherent Perfect Absorption phenomena has been demonstrated as a potential approach for WPT \cite{Oh2026}. 

We propose the FIO as an alternative, non-invasive method to maximally focus energy on a target embedded in a complex scattering environment, with advantages over those requiring phase conjugation, optimization, or time-reversal \cite{Prada1994,Taddese2009,Taddese2010,Ma2014,Zhou2014}. Unlike those methods, the FIO method does not require the target to emit guidestar-like signals, only to perturb the system's scattering matrix, and works even in systems with broken time-reversal symmetry due to high degrees of absorption or loss of reciprocity.

Other focusing techniques that work in non-reciprocal media include the use of the Wigner-Smith operator (WSO) \cite{Hougne2021} and the Generalized Wigner-Smith operator (GWSO) \cite{Ambichl2017,Orazbayev2024,Nerson2026}, which are closely related to the FIO. However, the WSO method requires that the target be the highest $Q$ object inside the system such that the long dwelling excitations are guaranteed to be at the target. The GWSO method is effectively identical as the FIO method for a unitary scattering matrix, but when considering a lossy system or sub-blocks of the scattering matrix that are generally sub-unitary, the FIO remains Hermitian while the GWSO does not. This means the GWSO does not have real, clearly ordered eigenvalues and eigenvectors that are guaranteed to be orthogonal and form a complete basis of all possible $|E_\mathrm{in}\rangle$ in the $M$ channel space. \rev{Additionally, the WSO and GWSO eigenvalues are maximized around scattering singularities where there is the least energy scattered back to the measuring receivers \cite{Erb2024,Shaibe_2025_Top}. Using WSO or GWSO methods therefore means working much closer to the noise floor, but the FIO eigenvalues are not correlated with scattering singularities (as we demonstrate in Appendix \ref{APP_SING}) so we do not suffer from that problem.}

\subsection{\rev{Measurement Scheme}}

Since the source of Fisher information is the parameter of interest \cite{Hupfl2024}, the incoming wave which returns the most information about the parameter should be the one that interacts with that parameter the most. Thus, the principal eigenvector of the FIO should be the most efficient excitation for depositing energy at the location of the parameter. We cannot directly measure the microwave intensity everywhere within the cavity, so we instead take advantage of the capability of the varactor metasurface to display non-linear behavior when stimulated with high incident power. We demonstrate that injecting the principal FIO eigenvector can drive our microwave system non-linear, whereas the other eigenvectors cannot. Note that the non-linearity is not being utilized as a guidestar, as is the case in some other methods \cite{Frazier2013,Goicoechea2025}, but is simply used as verification that the power has been sufficiently concentrated on the metasurface. Focusing using the FIO is of practical value for WPT applications because it simultaneously minimizes the field amplitudes elsewhere by construction.

For this set of measurements, we replace the quarter bowtie billiard with a circular billiard with a perimeter of $1.995$ m, shown in Fig.~\ref{FIG_3}(a). We place only one of the metasurfaces inside the circular billiard, so we can be certain that any non-linear behavior is solely due to this target. To target different locations inside the billiard, we can physically relocate the metasurface, which we demonstrate at two widely spaced locations, labeled A and B, which are further apart than the length of the metasurface. We measure the scattering matrix over both frequency and metasurface applied voltage, looking for a frequency-voltage pair $(\omega_0,V_0)$ where the largest magnitude ratio between any two elements of all four FIO eigenvectors is smaller than 20 dB. We do this because our \rev{internal (external)} sources have a limited dynamic range of \rev{$55$ dB ($39$ dB)}. Once $\omega_0$ and $V_0$ are fixed, we measure $|E_{\mathrm{out}}\rangle$ while injecting $c|E_{\mathrm{in}}\rangle$ where $c$ is a real scalar representing the power scaling and $|E_{\mathrm{in}}\rangle$ is an eigenvector of $F_V$.

\subsection{\rev{Evidence of Targeting}}

\begin{figure} [thb]
	\centering
    \includegraphics[width=1\columnwidth]{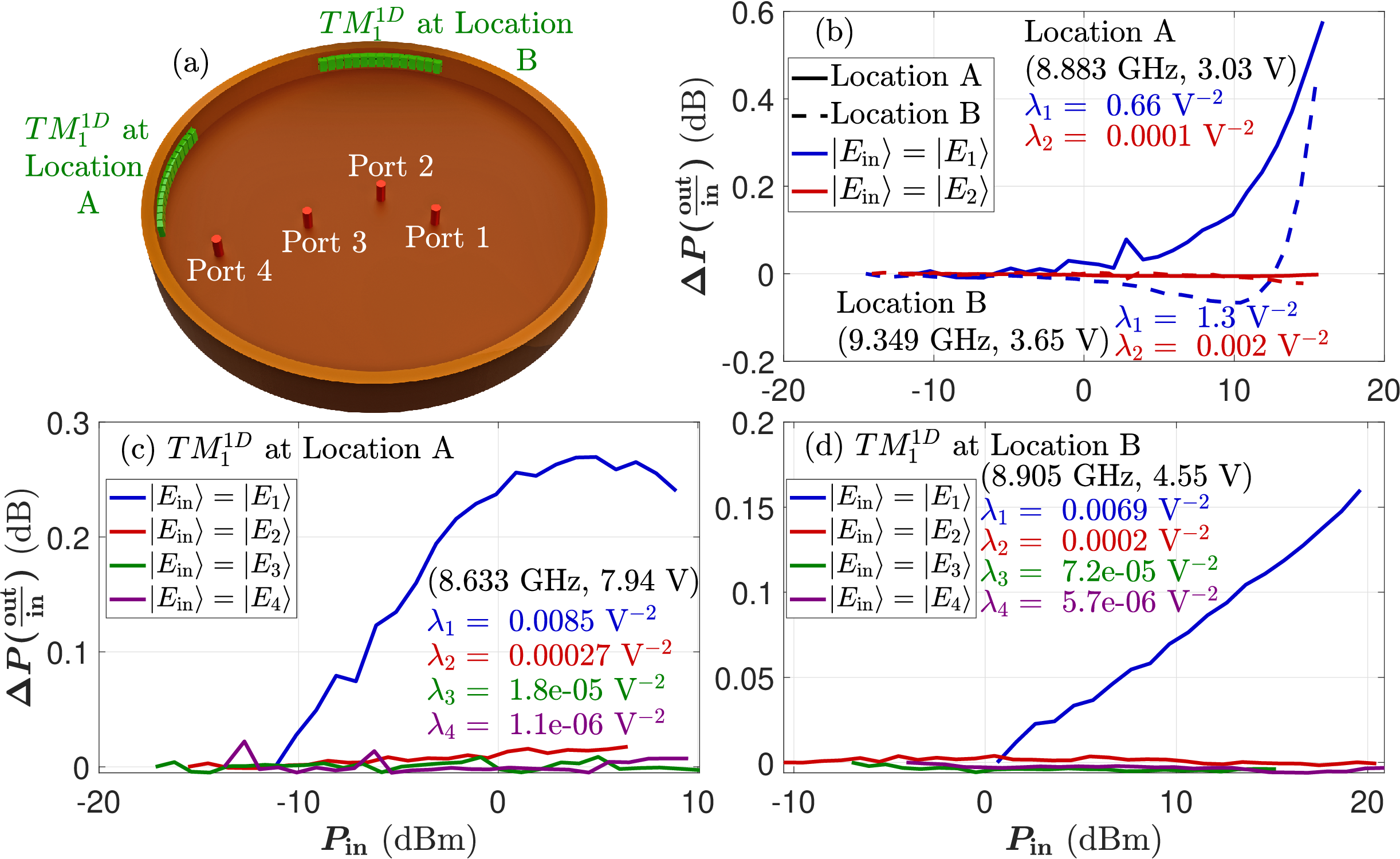} % One column
	\caption{(a) Schematic of circular microwave billiard (lid removed) with $M=4$ ports and a single metasurface $TM_1^{1D}$ which is moved between location A (on the left) and location B (at the top). (b) Change in the output-to-input power ratio $\Delta P(\frac{\mathrm{out}}{\mathrm{in}})$ with increasing total input power $P_\mathrm{in}$ in the case of $M=2$ ports. Solid (dashed) curves show $\Delta P(\frac{\mathrm{out}}{\mathrm{in}})$ when $TM^{1D}_1$ is placed in location A (B), $\omega_0$ and $V_0$ for each case \rev{is} given in the figure. (c-d) Change in the output-to-input power ratio $\Delta P(\frac{\mathrm{out}}{\mathrm{in}})$ with increasing input power $P_\mathrm{in}$ in the case of $M=4$ ports. $\Delta P(\frac{\mathrm{out}}{\mathrm{in}})$ is calculated as the change in the output-to-input power ratio from the smallest $P_\mathrm{in}$ measured for each $|E_{\mathrm{in}}\rangle$.}
	\label{FIG_3}
\end{figure}

While the scattering system remains linear, as the total input power $P_\mathrm{in}$ is increased, the output-to-input power ratio $P(\frac{\mathrm{out}}{\mathrm{in}})\equiv P_\mathrm{out}/P_\mathrm{in}$ remains constant. We therefore use a change in $P(\frac{\mathrm{out}}{\mathrm{in}})$ as an indication of non-linear behavior. In Fig.~\ref{FIG_3}(b), we plot the change in the ratio $P(\frac{\mathrm{out}}{\mathrm{in}})$  from the smallest input power $P_\mathrm{in}$ for four cases, and only two ports connected to the billiard. The solid blue (red) curve is $\Delta P(\frac{\mathrm{out}}{\mathrm{in}})$ when the metasurface is in location A, and $|E_\mathrm{in}\rangle=|E_1\rangle$ ($|E_\mathrm{in}\rangle=|E_2\rangle$). The blue curve shows exactly the behavior expected from a non-linear system while the red curve is completely flat at $\Delta P(\frac{\mathrm{out}}{\mathrm{in}})=0$ dB for all $P_\mathrm{in}$. This indicates that the principal FIO eigenvector does a good job of focusing energy on the metasurface and the second eigenvector does not. The dashed curves in Fig.~\ref{FIG_3}(b) show the same things but in the case of the metasurface at location B. It should be noted that in this case, the red curve does slightly deviate from $\Delta P(\frac{\mathrm{out}}{\mathrm{in}})=0$ dB at $P_\mathrm{in}\geq11$ dB, implying that the metasurface is being driven slightly non-linear despite the energy not being intentionally focused on the metasurface.

Panels (c-d) of Fig.~\ref{FIG_3} show the change in the output-to-input power ratio for the four port billiard. Because of the power dynamic range constraints on our microwave sources, we were not able to go low enough in $P_\mathrm{in}$ to actually see linear behavior for the  principal eigenvector excitation. This can be seen from the blue curves in both panels not having a flat region. Panel (c) in particular shows that non-linear behavior for $|E_\mathrm{in}\rangle=|E_1\rangle$ is seen all the way to $P_\mathrm{in}=-11$ dBm, however, we can likely attribute that to port 4 being so close to the metasurface when it is in location A.  For the sake of brevity, we only show data for two locations of the metasurfaces, and at one frequency for each. However, we are able to target the metasurface wherever it is in the billiard, and at any frequency at which the metasurface has a measurable impact on the scattering matrix. 
% THIS IS REDUNDANT: Demonstrating the change in $P(\frac{\mathrm{out}}{\mathrm{in}})$ is not possible in all cases because the dynamic ranges of the microwave sources constrains the range of power scaling $c$ of the eigenvector excitations.

We highlight three important considerations about this targeting protocol. First, we did not need to move the metasurface in order to target it, instead we changed its scattering properties by changing the voltage applied to the diodes. In this way, the method we propose here is very general, and any tunable or toggleable local perturber to the scattering matrix can be used as a target. The second is the corollary result, if it is desired to minimize the energy on the metasurface, we can do so by using the eigenvector of $F_V$ associated with its smallest eigenvalue as $|E_\mathrm{in}\rangle$. The third is that there are several orders of magnitude between the largest principal FIO eigenvalue shown ($\lambda_1=1.3~\mathrm{V}^{-2}$ in Fig.~\ref{FIG_3}(b)) and the smallest  ($\lambda_1=0.0069~\mathrm{V}^{-2}$ in Fig.~\ref{FIG_3}(d)), and yet focusing the energy onto the metasurface is possible in both cases. 
%\rev{The range of values of $\lambda_1$ poses an interesting question, and we provide a preliminary statistical investigation in Appendix~\ref{APP_STATS} to determine the likelihood having a large amount of Fisher information about the metasurface in the quarter bowtie billiard.}
%CUT FOR SPACE: This suggests that for practical applications of the Fisher information enabled techniques we describe, the principal eigenvalue of $F_x$ doesn't need to be maximized, just made ``large enough'', though at this moment we cannot say what ``large enough'' means in general as it would doubtless depend on the details of the parameter $x$.

\section{Conclusion}\label{SEC_Conc}

% \textit{Conclusion}.---
We have demonstrated how the FIO can be used to interrogate a complex scattering system and recover valuable information. Accurate quantitative estimates of small changes of a parameter $x$ can be made, either through just scattering matrix measurements or by directly exciting the system with the principal FIO eigenvector $|E_1\rangle$ using simultaneous multi-source injection. \rev{The precision of these estimates has been compared to the Cram\'{e}-Rao bound and found to be in good agreement}. A process of change estimation and counter perturbation can be done repeatedly over a period of time to keep the system stable at some desired state in what we call a ``discrete feedback loop''. The principal FIO eigenvector has also been shown to maximally focus the incoming energy on the spatial location of the varying parameter, which has potential applications in targeting and wireless power transfer. This method of focusing waves on a target in a complex medium has the advantages that it is non-invasive, non-iterative, and works even when the system is lossy or non-reciprocal, unlike other methods that rely on phase conjugation or time reversal. 

There are some limitations of the FIO technique, the largest being the derivative in $x$ required in Eqs.~\ref{EQN_EST}-\ref{EQN_FISH}. Some amount of control over the tuning of $x$ or knowledge of when it is switching states is needed to implement this method. Another major limitation of this protocol is that there is no way to use it to discriminate the source of a change in a scattering system.  All the changes of the system would be attributed to the sole parameter used to form the FIO, which leads to inaccurate estimates. We propose that a multi-parameter generalization of the FIO method might be possible through the use of a machine learning algorithm that could distinguish the cause of the perturbation  \cite{Ma2019}.

Finally,  we currently lack a quantitative way to determine from the value of  $\lambda_1$ whether enough ``information'' about the parameter can be retrieved when using $|E_1\rangle$ as the input excitation. Empirical experience seems to suggest $\lambda_1$ simply needs to be above a threshold, but that threshold depends on the system and parameter in question, without a theoretical way to predict what it should be in all cases. A future statistical investigation into the FIO eigenvalues with dependence on various system parameters such as number of channels $M$, presence or absence of symmetries, absorption strength, port coupling, etc., could help address this question, and the practicality of using the FIO in the ways laid out in this paper.

%TC:ignore

%\section{Acknowledgements}
\bigskip
\textbf{Acknowledgements} 
We thank David Shrekenhamer and Timothy Sleasman of JHU/APL for the design and fabrication of the metasurfaces used in this work. This work was partially supported by NSF/RINGS under grant No. ECCS-2148318, ONR under grant N000142312507, and DARPA WARDEN under grant HR00112120021.   

% \bigskip
% \textbf{Data Availability} 
% There are no publicly available research data or software supporting this manuscript. Requests for further information or data should be sent to the authors. 

%\clearpage
% \newpage 

%\renewcommand{\thefigure}{S\arabic{figure}}
\bigskip

%\textbf{MATERIALS AND METHODS} 

%\vspace{1cm}
\appendix

\section{\rev{Preliminary Investigation of FIO Eigenvalue Statistics}} \label{APP_STATS}

\rev{We present a preliminary investigation into the statistics of the FIO, focusing on the $M=4$ port quarter bowtie billiard and the varactor metasurface. All the conclusions we draw are strictly for this system and this parameter, and likely do not generalize to all cases. A more thorough and complete test which takes into account absorption, number and placement of channels, perturbation strength of the parameter, and other factors is still necessary.}

\rev{For this work, we measured the scattering matrix in an ergodic manner across frequency, applied bias voltage to the metasurface of interest, and applied bias voltage to the four other metasurfaces. Choosing $\Delta V=0.06~\mathrm{V}$, we determined $F_V$ for 40001 $\omega_0$'s between $9-10~\mathrm{GHz}$, 589 $V_0$'s between $0-6~\mathrm{V}$, and 256 cavity configurations using the four other metasurfaces. The total data used in this statistical study amounts to $1.36~\mathrm{TB}$.}

\begin{figure} [thb]
	\centering
    \includegraphics[width=1\columnwidth]{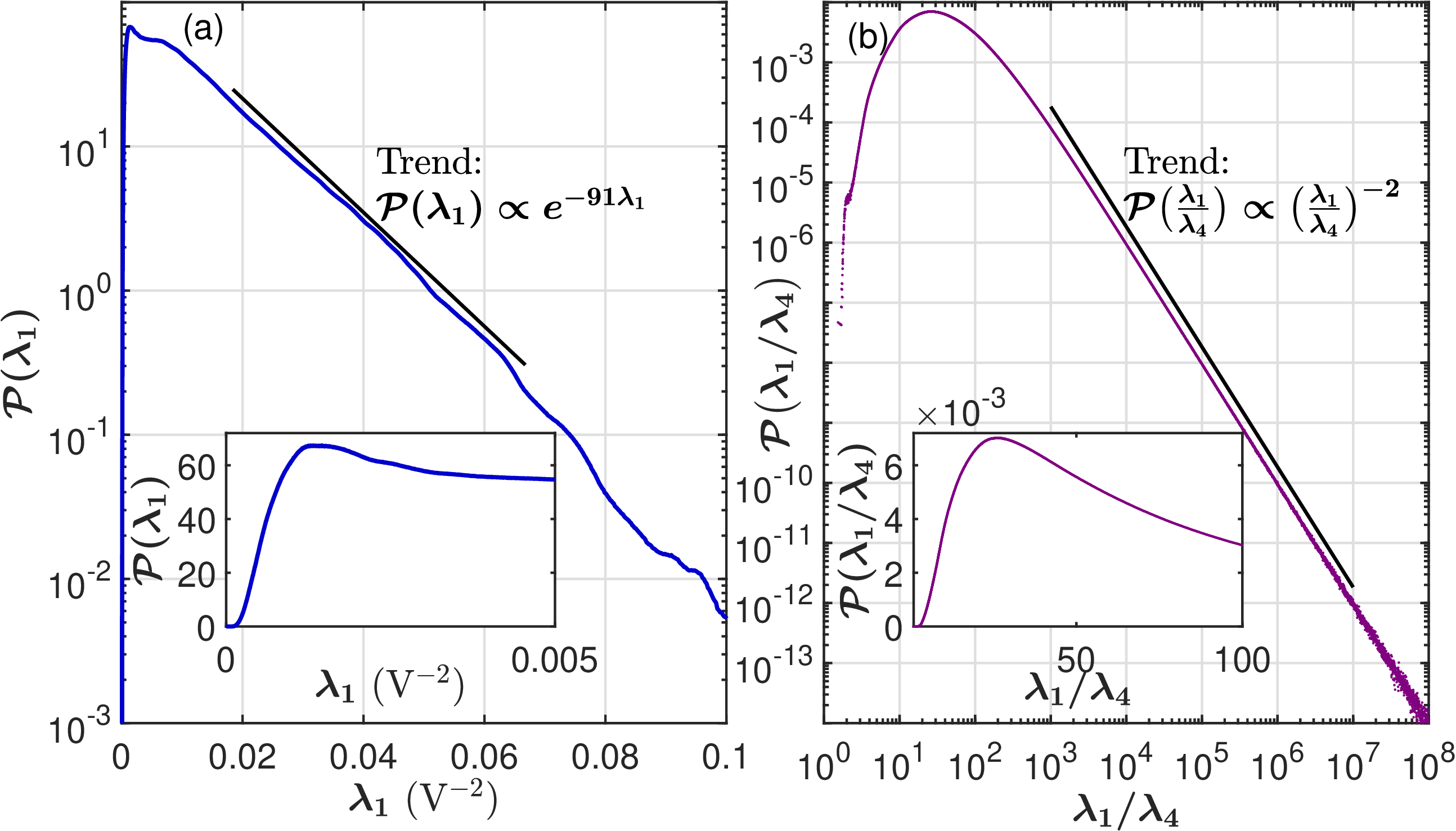} % One column
	\caption{\rev{(a) PDF of principal Fisher information operator eigenvalue $\lambda_1$ on a logarithmic-linear scale, black line characterizing the tail behavior as an exponential. (b) PDF of the ratio of the largest and smallest FIO eigenvalues $\lambda_1/\lambda_4$ on a log-log scale, black line characterizing the tail behavior as a $-2$ power law. Insets show the PDFs on a linear scale, zoomed in on the early and peak behavior. Data from the quarter-bow-tie billiard, as described in the text.}}
	\label{FIG_AppStats}
\end{figure}

\rev{In Fig.~\ref{FIG_AppStats}(a), we plot the PDF of $\lambda_1$ and find it linear on a logarithmic-linear scale. After performing a fit, we add the black line which characterizes an exponential with a decay constant of $-91$. Large values of $\lambda_1$ are therefore very strongly suppressed, but that does not mean the FIO is impractical, as we have shown successful estimations and targeting with $\lambda_1$'s much smaller than $0.1~\mathrm{V}^{-2}$. We note that the exponential decay of the Fisher information PDF may be a universal feature that is not unique to this system and parameter, as something similar was shown in the Supplementary Information of Ref.~\cite{Bouchet2021}. In the linear scale inset, we see the PDF has a flat zero valued foot in the limit of $\lambda_1\rightarrow0$ and what might be an exponential rise, suggesting a strong suppression with a lower bound on $\lambda_1$.}

\rev{We also consider the statistics of the dynamnic range of the FIO eigenvalues, and plot the PDF of the ratio $\lambda_1/\lambda_4$ in panel (b) on a log-log scale. For this ratio of eigenvalues, there is a $-2$ power-law tail that is clear over many decades, implying it's fairly common to have many orders of magnitude difference between the largest and smallest FIO eigenvalues. This tells us that different wavefronts have a tendency to experience the billiard very differently, and we should generically be able to switch between targeting and avoiding the metasurface through only wavefront shaping. The shape of the PDF also suggests that it is dominated by $\lambda_1$ at small values and by $1/\lambda_4$ at large values. The $-2$ power-law tail is indicative of Fisher information zeros being topologically stable lines in two-dimensional parameter space \cite{Shaibe_2025_Top}, but verification of this remains a future task.}

\section{\rev{Characterization of Noise in Measurements}} \label{APP_NOISE}

\begin{figure} [thb]
	\centering
    \includegraphics[width=1\columnwidth]{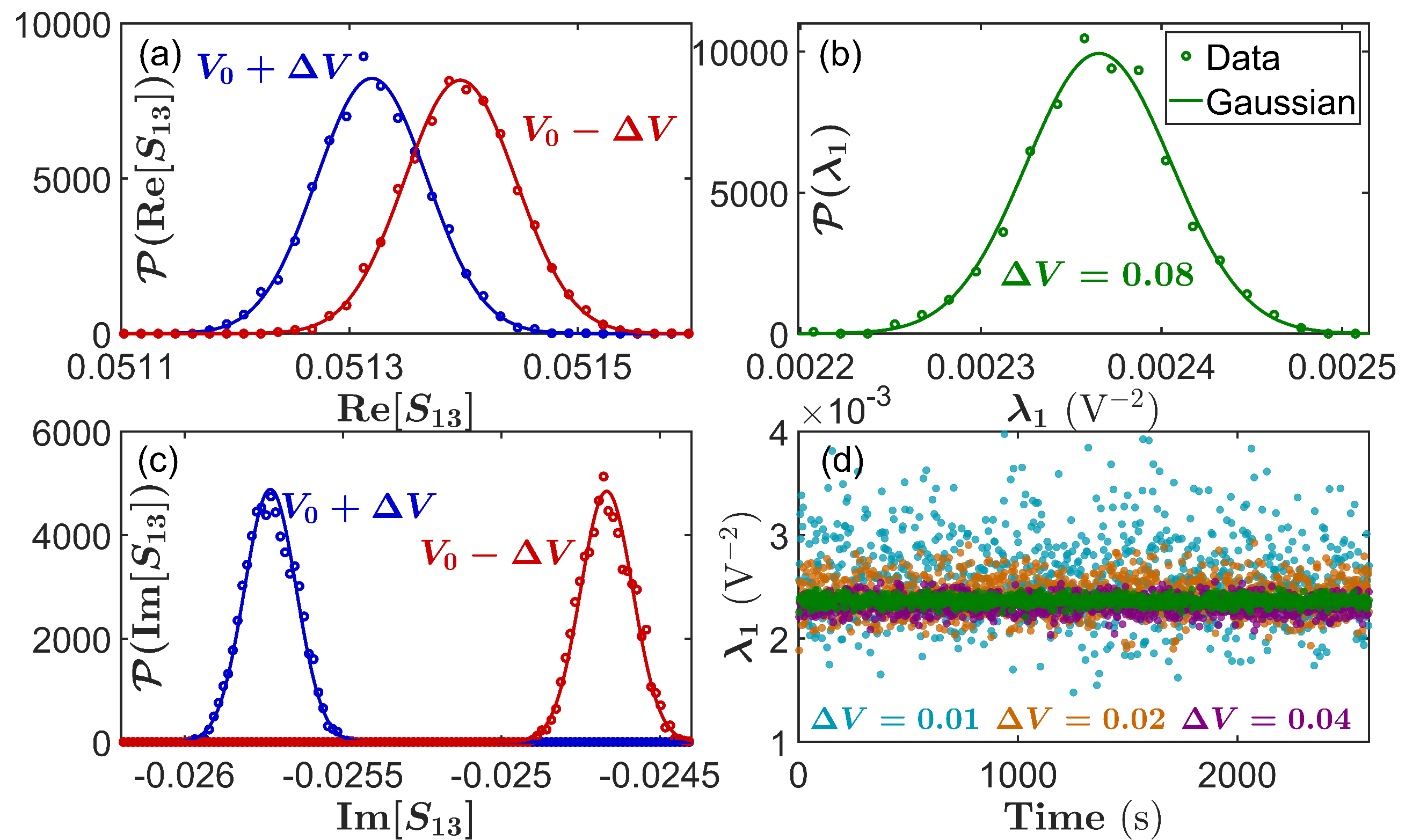} % One column
	\caption{\rev{Noise characterization measurement of the quarter bowtie billiard. The system was repeatedly measured at $8~\mathrm{GHz}$, $V_0=0.5~\mathrm{V}$, $\Delta V=0.08~\mathrm{V}$ for 3 hours. (a) and (c) show PDFs of $\mathrm{Re}[S_{13}]$ and $\mathrm{Im}[S_{13}]$, experimental data as circles and Gaussian fit as solid curves. (b) shows the PDF of the principal FIO eigenvalue $\lambda_1$. Panel (d) shows $\lambda_1$ over time for different choices of $\Delta V=[0.01,0.02,0.04,0.08]~\mathrm{V}$ (blue, orange, purple, green), and demonstrates that when $\lambda_1=0.0024~\mathrm{V}^{-2}$, there is not enough information to resolve small changes in the metasurface voltage from the background noise.}}
	\label{FIG_AppNoise}
\end{figure}

\rev{In this section, we characterize the noise in our measurements. In Ref.~\cite{Erb2026}, we characterized the stability of the billiard over time, and for the time scales relevant to this paper we do not expect the scattering properties to appreciably vary. The only sources of noise we expect are the digital instruments: the network analyzer and the power supply that provides the bias voltage to the metasurfaces, which should have Gaussian profiles as digital noise. To test this we repeatedly measure the scattering matrix of the $M=4$ port quarter bowtie billiard at $8~\mathrm{GHz}$ and $0.5\pm0.08~\mathrm{V}$ for three hours, resulting in 5000 measurements. We then show in Fig.~\ref{FIG_AppNoise}(a-c) the PDFs of $S_{13}$, an arbitrarily chosen element of $S$, and the principal eigenvalue $\lambda_1$ of $F_V$. Panels (a) and (c) have the PDFs of the real and imaginary parts of $S_{13}$, with the blue (red) being at $0.58~\mathrm{V}$ ($0.42~\mathrm{V}$). In panels (a)-(c), the circles are the experimental data and the solid curve is the Gaussian fit. In all cases shown, and for the scattering elements omitted for brevity, there is a very successful Gaussian fit to the data, allowing us to conclude the noise in our measurement is purely Gaussian.}

\rev{We performed follow up measurements where we did the same thing but used different values for $\Delta V$, and plot the resulting $\lambda_1$'s in panel (d). We see that the variance of $\lambda_1$ grows dramatically as $\Delta V$ shrinks, and the light blue dots corresponding to $\Delta V=0.01$ imply that we shouldn't be able to differentiate between states of the metasurface that are $\leq0.02~\mathrm{V}$ apart, something we were able to do in Fig.~\ref{FIG_2}(a) when we were implementing the discrete feedback loop. The reason for this discrepancy is that in this case $\lambda_1=0.0024~\mathrm{V}^{-2}$ while for Fig.~\ref{FIG_2}(a) we measured $\lambda_1=0.03~\mathrm{V}^{-2}$. This figure tell us that when there is less information, it is harder to make out small changes in the parameter and the noise dominates.}

\begin{figure} [thb]
	\centering
    \includegraphics[width=1\columnwidth]{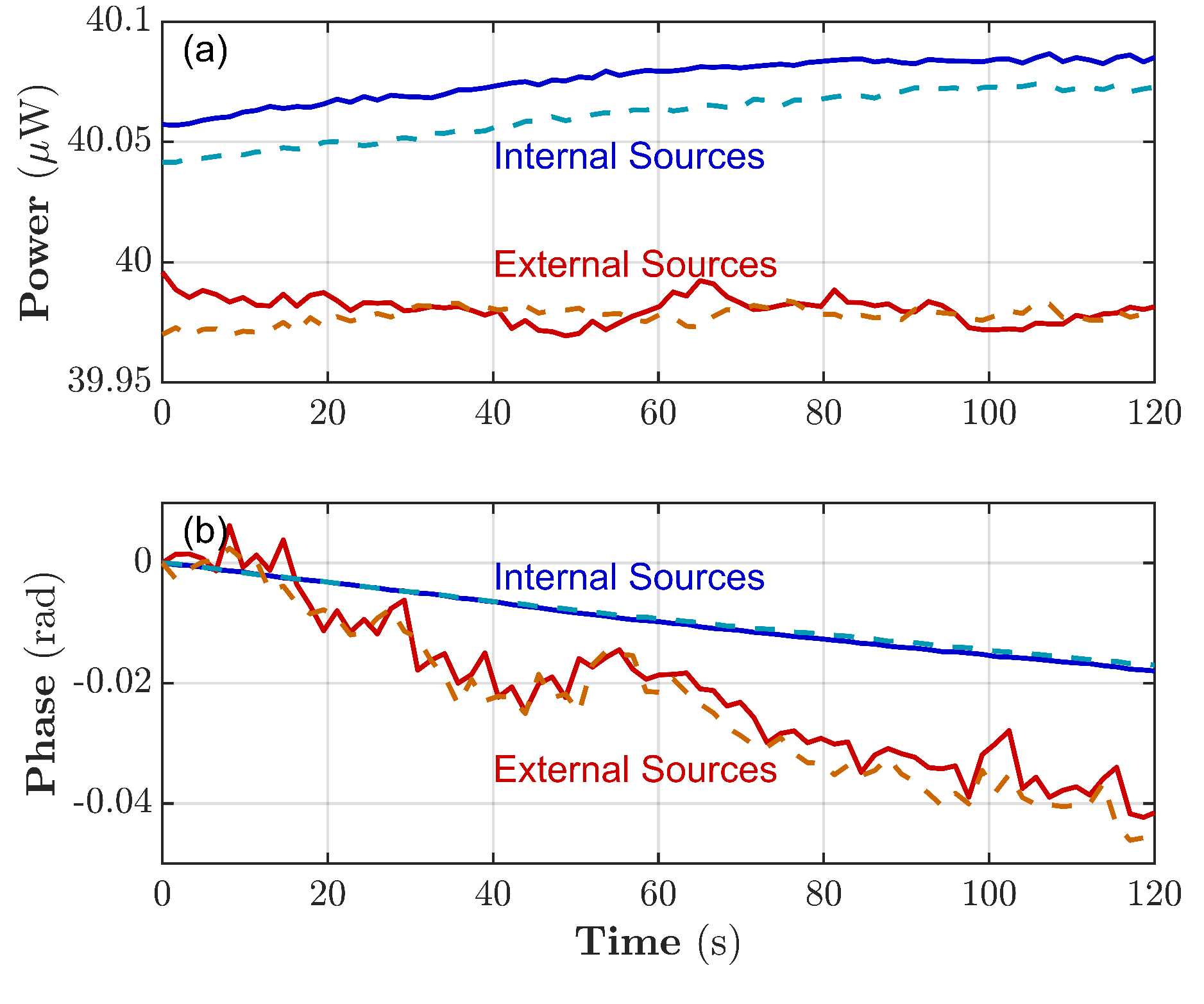} % One column
	\caption{\rev{Systematic drift of the microwave sources over 2 minutes. The network analyzer's internal sources are shown in blue and light blue while the two external sources are shown in red and orange. Panel (a) shows the power over time and panel (b) shows the phase over time.}}
	\label{FIG_AppSource}
\end{figure}

\rev{While not noise, there is also a systematic drift of the microwave sources when we do the multi-source injection measurements shown in Figs.~\ref{FIG_1},~\ref{FIG_2}(c), and \ref{FIG_3}. We categorize the four sources as the network analyzer's internal sources (blue and light blue) and the two external sources (red and orange). In Fig.~\ref{FIG_AppSource}, we show the time drift of these sources for both power and phase. While there is some drift in the internal sources, they remain mostly coherent over time, unlike the external sources which have powers and phases that intersect and separate. For the nearly instantaneous measurements shown in Fig.~\ref{FIG_1} or the measurements where $|E_{\mathrm{in}}\rangle$ has to be constantly changed as we reduce the total input power, the source drift is negligible. For the feedback loop measurement shown in Fig.~\ref{FIG_2}(c), the 101 iterations took 2.5 minutes to measure, and we used only the internal sources. Even the small amount of drift we see in Fig.~\ref{FIG_AppSource}, and specifically the slight decoherence in the phase, is enough to throw off our estimates of $\Delta V$ and cause a systematic error in the feedback loop. This demonstrates how sensitive the estimation is to $|E_\mathrm{in}\rangle.$}

\section{Fisher Information in Non-Reciprocal Scattering Systems} \label{APP_BTRI}

\begin{figure} [thb]
	\centering
	\includegraphics[width=1\columnwidth]{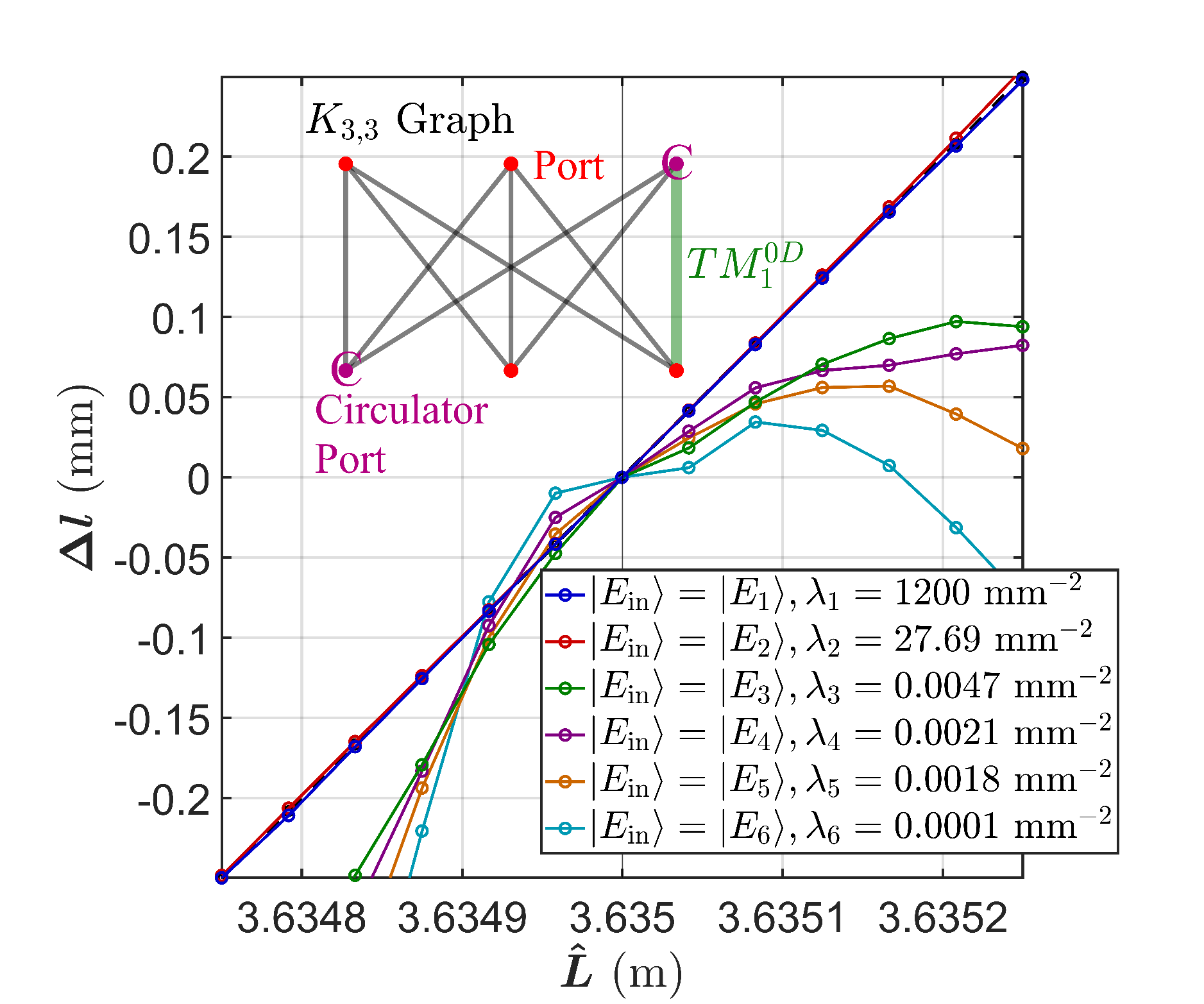}
	\caption{Estimate of the change in length $\Delta l$ of the phase shifter $TM_1^{0D}$ embedded along one bond of a six port, non-reciprocal microwave graph. Measurement conducted at $16$ GHz. Dashed black line depicts true change, which is a straight line with a slope of $1$. The associated eigenvalues of $F_L$ are given in the legend. Inset is a schematic of the $K_{3,3}$ microwave graph, with circulators creating non-reciprocal coupling marked in burgundy and the bond containing $TM_1^{0D}$ marked in green. Note that the symmetries of the bonds present in the schematic are not present in the physical graph, which utilizes bonds of incommensurate lengths.}
	\label{FIG_AppNR}
\end{figure}

In this section, we apply the FIO and Eq.~\ref{EQN_EST} to a non-reciprocal, lossy, irregular, microwave graph \cite{HulQG04} with the topology of the complete six node bipartite graph $K_{3,3}$, a schematic of which is shown as an inset in Fig.~\ref{FIG_AppNR}. All six nodes of the graph are connected to a six port network analyzer, and we add a three way microwave circulator to two of the nodes to introduce non-reciprocity. Because the circulators are on the port nodes and not ``interior nodes'' (which this graph lacks), it results in a special case of non-reciprocity which we call ``non-reciprocal coupling'' \cite{shaibe2025}. In addition, one of the bonds of the graph incorporates an analog microwave phase shifter $TM_1^{0D}$ to create a parametrically-tunable bond length.  The varying length changes the interference conditions at the nodes, thus altering the scattering matrix of the system.  The overall electrical length $L$ of the graph is $3.635\pm0.025$ m, where the $5$ cm is the tunable range of $TM_1^{0D}$.

We do not inject the six component eigenvectors directly into the graph to measure $|E_\mathrm{out}\rangle$ for this data, instead choosing to calculate it from the scattering matrix by $|E_\mathrm{out}(\hat{L})\rangle=S(\hat{L})|E_\mathrm{in}(L_0)\rangle$ for each value of $\hat{L}$ and each eigenvector of $F_L$ as a choice of $|E_\mathrm{in}\rangle$. We then use Eq.~\ref{EQN_EST} to estimate the change in length $\Delta l$ of $TM_1^{0D}$, which we can do accurately with $|E_\mathrm{in}\rangle=|E_1\rangle$ and $|E_\mathrm{in}\rangle=|E_2\rangle$ but not for the other eigenvectors, as shown in Fig.~\ref{FIG_AppNR}. 

We reiterate that $|E_1\rangle$ can be used as an ingoing wave to obtain good estimates of a parameter value change because it is the input excitation which maximizes the excitation amplitudes at the location of the varying parameter. Therefore showing accurate estimates of the change in length of $TM_1^{0D}$ implies the ability to focus the energy on that bond, and not the rest of the graph, by using $|E_1\rangle$ (and minimally excite that bond by using $|E_6\rangle$) as the ingoing wave.

\section{Fisher Information and Scattering Singularities} \label{APP_SING}

We find that the size of the FIO eigenvalues appears to be uncorrelated with scattering matrix singularities, which we define as a vortex of a complex scalar field derived from the scattering matrix \cite{Shaibe_2025_Top}. Even though the scattering matrix changes most dramatically with any parameter around scattering singularities \cite{Erb2024}, singularities are not where the scattering matrix carries the most information about the change of a particular parameter. We demonstrate this in the case of the two port tetrahedral graph described in Ref.~\cite{Shaibe_2025_Top}.

\begin{figure} [thb]
	\centering
	\includegraphics[width=1\columnwidth]{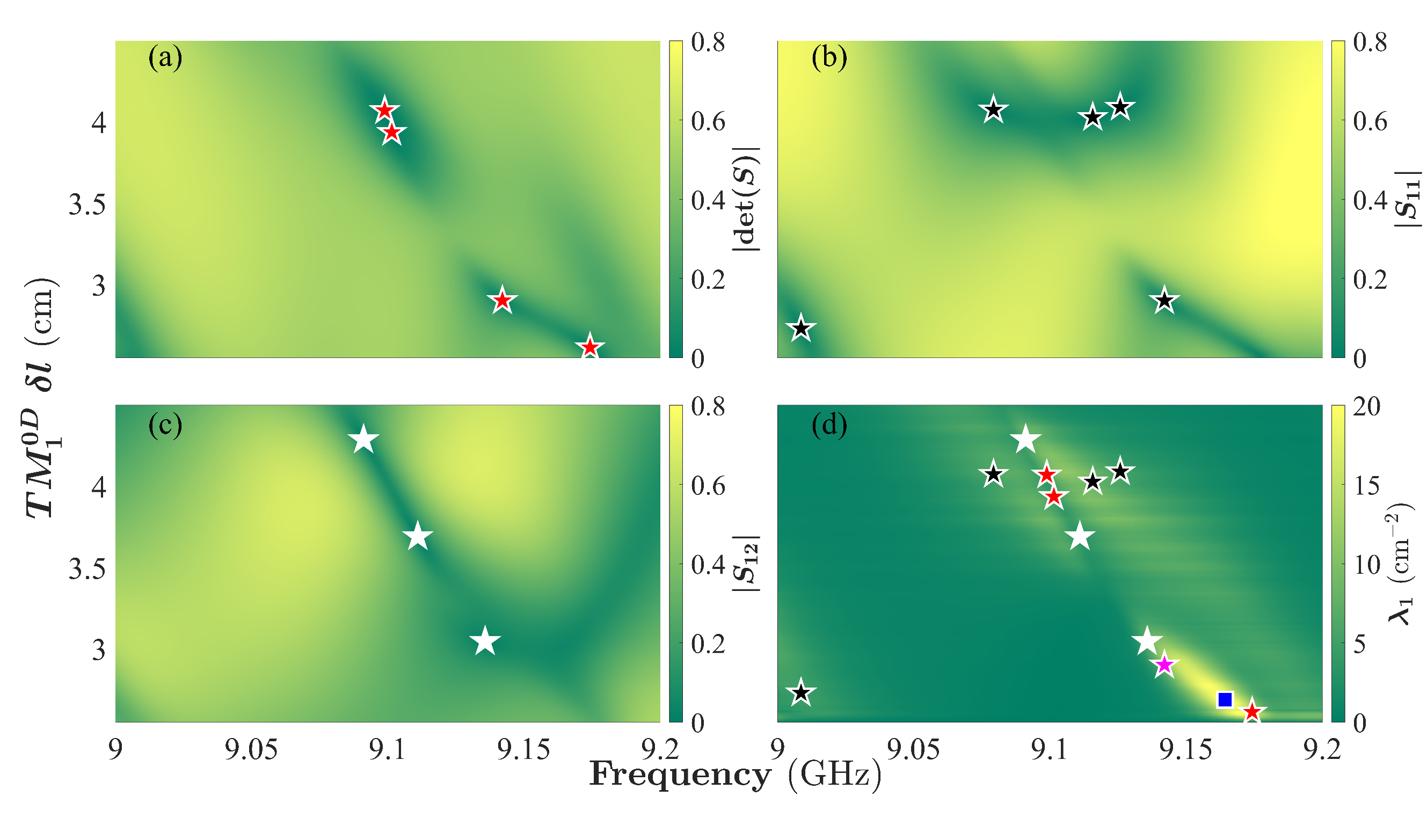}
	\caption{Experimental scattering matrix data from two port tetrahedral microwave graph measured in the two-parameter space of frequency and length change $\delta l$ of phase shifter $TM_1^{0D}$. (a) Magnitude of $\mathrm{det}(S)$, red stars highlight points of $\mathrm{det}(S)=0+i0$ which is the coherent perfect absorption (CPA) enabling condition. (b) Magnitude of $S_{11}$, black stars highlight points of $S_{11}=0+i0$. (c) Magnitude of $S_{12}$, white stars highlight points of $S_{12}=0+i0$. Note that due to reciprocity, $S_{12}=S_{21}$. (d) Principal eigenvalue $\lambda_1$ of the Fisher information operator $F_{\delta l}$. Symbols carried over from panels (a-c), with blue square marking the largest value of $\lambda_1$ in the parameter space and the pink star representing the coincident $\mathrm{det}(S)=0+i0$ and $S_{11}=0+i0$.}
	\label{FIG_AppSing}
\end{figure}

This graph is constructed using the same phase shifter $TM^{0D}_1$ as the six node graph discussed in Appendix \ref{APP_BTRI}. We measure the scattering matrix of a 2-port tetrahedral graph in the two dimensional parameter space of frequency and bond-length change $\delta l$ of $TM^{0D}_1$. We locate all the scattering singularities of $\mathrm{det}(S)$, $S_{11}$, and $S_{12}$ in this parameter space, and plot their speckle patterns in Fig.~\ref{FIG_AppSing}(a-c). We identify the scattering singularities by the presence of topological phase windings \cite{Berry2000,Dennis2025,erb2025}. We then form $F_{\delta l}$ and plot its principal eigenvalue $\lambda_1$ in panel (d) of Fig.~\ref{FIG_AppSing}. The value of $\lambda_1$ does not peak at any of these singularities, or $S_{22}=0+i0$ singularities which are omitted for \rev{clarity in the figure}. This finding proves scattering singularities cannot, in general, be used to maximize the value of the maximum Fisher eigenvalue $\lambda_1$. This is in contrast to complex time delay (related to the eigenvalues of the WSO) \cite{Wigner1955,Smith1960,Asano2016,Lei2021,Kang2021,Huang2022,Erb2024,shaibe2025,Shaibe_2025_Top} or complex generalized response (related to the eigenvalues of the GWSO) \cite{Shaibe_2025_Top} which diverge at scattering singularities.

%\section{\label{sec:Fig}Figures}  

% \newpage

% \textbf{EXTENDED DATA} 
%\vspace{1cm}

\clearpage
\newpage

%TC:endignore
\bibliography{Refs.bib}

%\bigskip
%\textbf{Author contributions} N.S. performed measurements, analysis, and wrote the initial manuscript. S.M.A. conceived and directed the research. J.E. performed measurements.

%\bigskip
%\textbf{Competing interests}  The authors have no competing interests to declare. 

%\bigskip
%\textbf{Data and Materials Availability}  All data are available in the main text or the supplementary materials. 

%\bigskip
%\textbf{Correspondence}  and requests for materials should be addressed to Nadav Shaibe.

%\bigskip
%\textbf{Suggested Referees}  

\end{document}